\documentstyle[amssymb,12pt]{article}

\if@twoside
\else
\fi
\advance\textheight by \topskip
\leftmargin\leftmargini
\labelwidth\leftmargini\advance\labelwidth-\labelsep

\catcode`\@=11
\def\Let@{\relax\iffalse{\fi\let\\=\cr\iffalse}\fi}
\def\vspace@{\def\vspace##1{\crcr\noalign{\vskip##1\relax}}}
\def\multilimits@{\bgroup\vspace@\Let@
 \baselineskip\fontdimen10 \scriptfont\tw@
 \advance\baselineskip\fontdimen12 \scriptfont\tw@
 \lineskip\thr@@\fontdimen8 \scriptfont\thr@@
 \lineskiplimit\lineskip
 \vbox\bgroup\ialign\bgroup\hfil$\m@th\scriptstyle{##}$\hfil\crcr}
\def\Sb{_\multilimits@}
\def\endSb{\crcr\egroup\egroup\egroup}
\def\Sp{^\multilimits@}

\long
\def\QQQ#1#2{}
\def\QTP#1{}
\long
\def\QQA#1#2{}

\def\EXPAND#1[#2]#3{}
\def\NOEXPAND#1[#2]#3{}

\def\LaTeXparent#1{}

\newdimen\ex@
\def\rightarrowfill@#1{$#1\m@th\mathord-\mkern-6mu\cleaders
 \hbox{$#1\mkern-2mu\mathord-\mkern-2mu$}\hfill
 \mkern-6mu\mathord\rightarrow$}%
\def\leftarrowfill@#1{$#1\m@th\mathord\leftarrow\mkern-6mu\cleaders
 \hbox{$#1\mkern-2mu\mathord-\mkern-2mu$}\hfill\mkern-6mu\mathord-$}%
\def\leftrightarrowfill@#1{$#1\m@th\mathord\leftarrow
\mkern-6mu\cleaders
 \hbox{$#1\mkern-2mu\mathord-\mkern-2mu$}\hfill
 \mkern-6mu\mathord\rightarrow$}%
\def\overrightarrow{\mathpalette\overrightarrow@}%
\def\overrightarrow@#1#2{\vbox{\ialign{##\crcr\rightarrowfill@#1\crcr
 \noalign{\kern-\ex@\nointerlineskip}$\m@th\hfil#1#2\hfil$\crcr}}}%

\def\overleftarrow{\mathpalette\overleftarrow@}%
\def\overleftarrow@#1#2{\vbox{\ialign{##\crcr\leftarrowfill@#1\crcr
 \noalign{\kern-\ex@\nointerlineskip}$\m@th\hfil#1#2\hfil$\crcr}}}%
\def\overleftrightarrow{\mathpalette\overleftrightarrow@}%
\def\overleftrightarrow@#1#2{\vbox{\ialign{##\crcr
   \leftrightarrowfill@#1\crcr
 \noalign{\kern-\ex@\nointerlineskip}$\m@th\hfil#1#2\hfil$\crcr}}}%
\def\underrightarrow{\mathpalette\underrightarrow@}%
\def\underrightarrow@#1#2{\vtop{\ialign{##\crcr$\m@th\hfil#1#2\hfil
  $\crcr\noalign{\nointerlineskip}\rightarrowfill@#1\crcr}}}%

\def\underleftarrow{\mathpalette\underleftarrow@}%
\def\underleftarrow@#1#2{\vtop{\ialign{##\crcr$\m@th\hfil#1#2\hfil
  $\crcr\noalign{\nointerlineskip}\leftarrowfill@#1\crcr}}}%
\def\underleftrightarrow{\mathpalette\underleftrightarrow@}%
\def\underleftrightarrow@#1#2{\vtop{\ialign{##\crcr$\m@th
  \hfil#1#2\hfil$\crcr
 \noalign{\nointerlineskip}\leftrightarrowfill@#1\crcr}}}%

\def\RIfM@{\relax\ifmmode}

\newif\iffirstchoice@
\firstchoice@true
\def\textfonti{\the\textfont\@ne}
\def\textfontii{\the\textfont\tw@}
\def\text{\RIfM@\expandafter\text@\else\expandafter\text@@\fi}
\def\text@@#1{\leavevmode\hbox{#1}}
\def\text@#1{\mathchoice
 {\hbox{\everymath{\displaystyle}\def\textfonti{\the\textfont\@ne}%
  \def\textfontii{\the\textfont\tw@}\textdef@@ T#1}}
 {\hbox{\firstchoice@false
  \everymath{\textstyle}\def\textfonti{\the\textfont\@ne}%
  \def\textfontii{\the\textfont\tw@}\textdef@@ T#1}}
 {\hbox{\firstchoice@false
  \everymath{\scriptstyle}\def\textfonti{\the\scriptfont\@ne}%
  \def\textfontii{\the\scriptfont\tw@}\textdef@@ S\rm#1}}
 {\hbox{\firstchoice@false
  \everymath{\scriptscriptstyle}\def\textfonti
  {\the\scriptscriptfont\@ne}%
  \def\textfontii{\the\scriptscriptfont\tw@}\textdef@@ s\rm#1}}}
\def\textdef@@#1{\textdef@#1\rm\textdef@#1\bf\textdef@#1\sl\textdef@#1\it}
\def\DN@{\def\next@}
\def\eat@#1{}
\def\textdef@#1#2{%
 \DN@{\csname\expandafter\eat@\string#2fam\endcsname}%
 \if S#1\edef#2{\the\scriptfont\next@\relax}%
 \else\if s#1\edef#2{\the\scriptscriptfont\next@\relax}%
 \else\edef#2{\the\textfont\next@\relax}\fi\fi}

\QQQ{Language}{
American English
}

\begin{document}

\vspace*{1cm}

\begin{center}
{\LARGE Remarks on Duffin-Kemmer-Petiau theory and gauge invariance}
\end{center}

\vspace{1cm}

\begin{center}
{\large J. T. Lunardi}\footnote{%
On leave from Departamento de Matem\'{a}tica e Estat\'{\i}stica. Setor de
Ci\^{e}ncias Exatas e Naturais. Universidade Estadual de Ponta Grossa. Ponta
Grossa, PR - Brazil.}{\large , B. M. Pimentel} {\large , R. G. Teixeira} 
{\large and J. S. Valverde}

\vskip 0.5cm

Instituto de F\'{\i}sica Te\'{o}rica

Universidade Estadual Paulista

Rua Pamplona 145

01405-900 - S\~{a}o Paulo, S.P.

Brazil
\end{center}

\vskip 1cm

\begin{center}
\begin{minipage}{14.5cm}
\centerline{\bf Abstract}

{Two problems relative to the electromagnetic coupling of Duffin-Kemmer-Petiau (DKP)
theory are discussed: the presence of an {\it anomalous} term in the Hamiltonian form of
the theory and the apparent difference between the Interaction terms in DKP 
and Klein-Gordon (KG) Lagrangians. For this, we first discuss the behavior of DKP field 
and its physical components under gauge transformations. From this analysis, we can 
show that these problems simply do not exist if one correctly analyses the physical 
components of DKP field.}
\end{minipage}
\end{center}

\newpage

\section{Introduction}

The Duffin-Kemmer-Petiau (DKP) equation is a first order relativistic wave
equation for spin 0 and 1 bosons \cite{Duffin, Kemmer, Petiau}, similar to
Dirac equation. The historical development of this theory, among others,
until the 70's can be found in reference \cite{Krajcik}. More recently there
have been an increasing interest in DKP theory, specifically it has been
applied to QCD (large and short distances)\ by Gribov \cite{Gribov}, to
covariant Hamiltonian dinamics by Kanatchikov \cite{Kanatchikov} and have
been generalized to curved space-time by Red'kov \cite{Red'kov} and Lunardi 
{\it et al} \cite{Lunardi}.

It is well known that for free fields there is a perfect equivalence between
DKP equation and Klein-Gordon (KG) and Proca equations but, when interaction
with electromagnetic field through minimal coupling is present, doubts about
this equivalence arise. This is because when one pass to the Hamiltonian
form (Schr\"{o}dinger like) or to a second order KG like wave equation an 
{\it anomalous} term that does not posses a clear physical interpretation
appears \cite{Kemmer}. Moreover, it was argued that, when applied to the
Lagrangian density for the spin 0 version of DKP theory, the minimal
coupling seems to provide an interaction term linear in the potential vector 
$A^{\mu }$, a different result from the quadratic term in $A^{\mu }$ found
when starting with the KG Lagrangian.

Our intention in this paper is to show that the problem in the physical
interpretation of DKP theory and the contradiction between it and KG theory
are only apparent and can be fully elucidated by a correct interpretation of
the physical meaning of the DKP wave function $\psi $ and the correct
application of gauge invariance principle.

For this, we will dedicate Section 2 to mention some basic results in DKP
free field theory, mainly those relative to the obtainment and
interpretation of the physical components of DKP field. For further details
we suggest the reader to original works \cite{Duffin, Kemmer} or classic
textbooks \cite{Umezawa, Berestetskii}. Next, in Section 3, we analyse the
apparent discrepancy between KG and DKP theories when electromagnetic
interaction comes to play. This discrepancy is shown to be caused by the use
of an expression for the DKP field, in terms of its physical components,
that is incompatible with the gauge invariance. Once we construct a correct
expression for the DKP field we find that there is a perfect agreement.

In Section 4 we show that the question of the apparently problematic term
lacking physical interpretation can be shown to be not relevant by noticing
that it disappears when we analyse the physical components of the DKP field $%
\psi $. This fact has already been shown by Nowakowski \cite{Nowakowski} at
classical level\ and more recently by Fainberg and Pimentel \cite{Fainberg}
at classical and quantum level too. But in this paper we adopt a different
approach to reduce the wave function $\psi $ to its physical components
using the projectors on the spin 0 and spin 1 sectors of DKP theory \cite
{Umezawa} in order to make more clear, in our opinion, the origin of this
physical interpretation. Finally, in Section 5, we make some concluding
remarks.

\section{The free field theory}

The DKP equation is given by 
\begin{equation}
\left( i\beta ^{\mu }\partial _{\mu }-m\right) \psi =0,  \label{eq1}
\end{equation}
where the matrices $\beta ^{\mu }$ obey the algebraic relations 
\begin{equation}
\beta ^{\mu }\beta ^{\nu }\beta ^{\rho }+\beta ^{\rho }\beta ^{\nu }\beta
^{\mu }=\beta ^{\mu }\eta ^{\nu \rho }+\beta ^{\rho }\eta ^{\nu \mu },
\label{eq2}
\end{equation}
being $\eta ^{\mu \nu }$ the metric tensor of Minkowski space-time with
signature $\left( +---\right) $. The matrices $\beta ^{\mu }$ have 2 non
trivial irreducible representations: a 5 degree one, corresponding to spin 0
particles, and a 10 degree one, corresponding to spin 1 particles. From the
algebraic relation above we can also obtain (no summation on repeated
indexes) 
\begin{equation}
\left( \beta ^{\mu }\right) ^{3}=\eta ^{\mu \mu }\beta ^{\mu },  \label{eq3}
\end{equation}
so that we can define the matrices 
\begin{equation}
\eta ^{\mu }=2\left( \beta ^{\mu }\right) ^{2}-\eta ^{\mu \mu }\ \left( 
\text{no summation}\right)  \label{eq3b}
\end{equation}
that satisfy 
\begin{equation}
\left( \eta ^{\mu }\right) ^{2}=1,\ \eta ^{\mu }\eta ^{\nu }-\eta ^{\nu
}\eta ^{\mu }=0,  \label{eq3c}
\end{equation}
\begin{equation}
\eta ^{\mu }\beta ^{\nu }+\beta ^{\nu }\eta ^{\mu }=0\ \left( \mu \neq \nu
\right) ,  \label{eq4}
\end{equation}
\begin{equation}
\eta ^{\mu \mu }\beta ^{\mu }=\eta ^{\mu }\beta ^{\mu }=\beta ^{\mu }\eta
^{\mu }\ \left( \text{no summation}\right) .  \label{eq5}
\end{equation}

With these results we can write the Lagrangian density for DKP free field as 
\begin{equation}
{\cal L}=\frac{i}{2}\overline{\psi }\beta ^{\mu }\overleftrightarrow{%
\partial }_{\mu }\psi -m\overline{\psi }\psi ,  \label{eq6}
\end{equation}
where $\overline{\psi }$ is defined\ as 
\begin{equation}
\overline{\psi }=\psi ^{\dagger }\eta ^{0}.  \label{eq6b}
\end{equation}
Moreover, we can choose $\beta ^{0}$ to be hermitian and $\beta ^{i}$
anti-hermitian so that the equation for $\overline{\psi }$ can also be
easily obtained by applying hermitian conjugation to equation (\ref{eq1}).

Multiplying the DKP equation by $\partial _{\alpha }\beta ^{\alpha }\beta
^{\nu }$ from left we can find \cite{Kemmer}

\begin{equation}
\partial ^{\nu }\psi =\beta ^{\alpha }\beta ^{\nu }\partial _{\alpha }\psi
\label{eq6c}
\end{equation}
and contracting this result with $\partial _{\nu }$ we have 
\begin{equation}
\square \psi +m^{2}\psi =0  \label{eq6d}
\end{equation}
so that each component of $\psi $ satisfies KG equation, as expected for a
relativistic field equation. Summing equation (\ref{eq6c}) for $\nu =0$ and
DKP equation multiplied from left by $-i\beta ^{0}$ we get a Schr\"{o}dinger
like Hamiltonian form wave equation 
\begin{equation}
i\partial _{t}\psi =H\psi  \label{eq6e}
\end{equation}
where 
\begin{equation}
H=i\left[ \beta ^{i},\beta ^{0}\right] \partial _{i}+m\beta ^{0}.
\label{eq6f}
\end{equation}

Under a Lorentz transformation $x^{\prime \mu }=\Lambda ^{\mu }{}_{\nu
}x^{\nu }$ we have 
\begin{equation}
\psi \rightarrow \psi ^{\prime }=U\left( \Lambda \right) \psi ,  \label{eq7}
\end{equation}
\begin{equation}
U^{-1}\beta ^{\mu }U=\Lambda ^{\mu }{}_{\nu }\beta ^{\nu },
\end{equation}
and for infinitesimal transformations $\Lambda ^{\mu \nu }=\eta ^{\mu \nu
}+\omega ^{\mu \nu }$ $\left( \omega ^{\mu \nu }=-\omega ^{\nu \mu }\right) $
we obtain \cite{Umezawa} 
\begin{equation}
U=1+\frac{1}{2}\omega ^{\mu \nu }S_{\mu \nu },\ S_{\mu \nu }=\left[ \beta
_{\mu },\beta _{\nu }\right] .  \label{eq9}
\end{equation}

The spin 0 and spin 1 sectors of the theory can be selected from a general
representation of $\beta ^{\mu }$ matrices through a set of operators, as
shown in reference \cite{Umezawa}. For the spin 0 sector the operators are 
\begin{equation}
P=-\left( \beta ^{0}\right) ^{2}\left( \beta ^{1}\right) ^{2}\left( \beta
^{2}\right) ^{2}\left( \beta ^{3}\right) ^{2},  \label{eq11}
\end{equation}
which satisfies $P^{2}=P$, and 
\begin{equation}
P^{\mu }=P\beta ^{\mu }.  \label{eq12}
\end{equation}

It can be shown that 
\begin{equation}
P^{\mu }\beta ^{\nu }=P\eta ^{\mu \nu },\ PS_{\mu \nu }=0,  \label{eq13}
\end{equation}
and, as consequence, under infinitesimal Lorentz transformations (\ref{eq9})
we have 
\begin{equation}
PU\psi =P\psi ,
\end{equation}
so that $P\psi $ transforms as a (pseudo)scalar. Similarly 
\begin{equation}
P^{\mu }U\psi =P^{\mu }\psi +\omega ^{\mu }{}_{\nu }P^{\nu }\psi ,
\label{eq15}
\end{equation}
showing that $P^{\mu }\psi $ transforms like a (pseudo)vector.

Applying these operators to DKP equation (\ref{eq1}) we have 
\begin{equation}
\partial _{\nu }\left( P^{\nu }\psi \right) =\frac{m}{i}P\psi ,  \label{eq16}
\end{equation}
and 
\begin{equation}
P^{\nu }\psi =\frac{i}{m}\partial ^{\nu }\left( P\psi \right) ,  \label{eq17}
\end{equation}
which combined provide 
\begin{equation}
\partial ^{\mu }\partial _{\mu }\left( P\psi \right) +m^{2}\left( P\psi
\right) =\square \left( P\psi \right) +m^{2}\left( P\psi \right) =0.
\label{eq18}
\end{equation}

These results show that all elements of the column matrix $P\psi $ are
scalar fields of mass $m$ obeying KG equation while the elements of $P^{\mu
}\psi $ are $\frac{i}{m}$ times the derivative with respect to $x_{\mu }$ of
the corresponding elements of $P\psi $. Then, acting $P$ upon $\psi $
selects the spin 0 sector of DKP theory, making explicitly clear that it
describes a scalar particle.

Now, for the spin 1 sector we have as operators 
\begin{equation}
R^{\mu }=\left\{ 
\begin{array}{c}
\left( \beta ^{1}\right) ^{2}\left( \beta ^{2}\right) ^{2}\left( \beta
^{3}\right) ^{2}\beta ^{\mu }\beta ^{0};\ \mu \neq 0 \\ 
-\left( \beta ^{1}\right) ^{2}\left( \beta ^{2}\right) ^{2}\left( \beta
^{3}\right) ^{2}\left( 1-\left( \beta ^{0}\right) ^{2}\right) ;\ \mu =0
\end{array}
\right. ;  \label{eq18-1}
\end{equation}
which can be written compactly as 
\begin{equation}
R^{\mu }=\left( \beta ^{1}\right) ^{2}\left( \beta ^{2}\right) ^{2}\left(
\beta ^{3}\right) ^{2}\left[ \beta ^{\mu }\beta ^{0}-\eta ^{\mu 0}\right] ;
\label{eq18-1b}
\end{equation}
and 
\begin{equation}
R^{\mu \nu }=R^{\mu }\beta ^{\nu }.  \label{eq18-2}
\end{equation}

From these definitions we have the following properties 
\begin{equation}
R^{\mu \nu }=-R^{\nu \mu },  \label{eq18-3}
\end{equation}
\begin{equation}
R^{\mu }\beta ^{\nu }\beta ^{\alpha }=\eta ^{\nu \alpha }R^{\mu }-\eta ^{\mu
\alpha }R^{\nu },  \label{eq18-4}
\end{equation}
\begin{equation}
R^{\mu }S^{\nu \alpha }=\eta ^{\mu \nu }R^{\alpha }-\eta ^{\mu \alpha
}R^{\nu },  \label{eq18-5}
\end{equation}
\begin{equation}
R^{\mu \nu }S^{\alpha \beta }=\eta ^{\nu \alpha }R^{\mu \beta }-\eta ^{\mu
\alpha }R^{\nu \beta }-\eta ^{\nu \beta }R^{\mu \alpha }+\eta ^{\mu \beta
}R^{\nu \alpha }.  \label{eq18-6}
\end{equation}

These results allow us to show that under infinitesimal Lorentz
transformations we have 
\begin{equation}
R^{\mu }U\psi =R^{\mu }\psi +\omega ^{\mu }{}_{\alpha }R^{\alpha }\psi ,
\label{eq18-7}
\end{equation}
so we can see that $R^{\mu }\psi $ transforms like a (pseudo)vector while $%
R^{\mu \nu }\psi $ transforms like a (pseudo)tensor since 
\begin{equation}
R^{\mu \nu }U\psi =R^{\mu \nu }\psi +\omega ^{\nu }{}_{\beta }R^{\mu \beta
}\psi +\omega ^{\mu }{}_{\alpha }R^{\alpha \nu }\psi .  \label{eq18-8}
\end{equation}

The application of these operators to DKP equation results in 
\begin{equation}
\partial _{\nu }\left( R^{\mu \nu }\psi \right) =\frac{m}{i}R^{\mu }\psi ,
\label{eq18-9}
\end{equation}
and 
\begin{equation}
R^{\mu \alpha }\psi =-\frac{i}{m}U^{\mu \alpha },  \label{eq18-10}
\end{equation}
where 
\begin{equation}
U^{\mu \alpha }=\partial ^{\mu }R^{\alpha }\psi -\partial ^{\alpha }R^{\mu
}\psi  \label{eq18-11}
\end{equation}
is the strength tensor of the massive vector field $R^{\mu }\psi $.
Combined, these results provide 
\begin{equation}
\partial _{\nu }\left( -\frac{i}{m}U^{\mu \nu }\right) =\frac{m}{i}R^{\mu
}\psi  \label{eq18-11b}
\end{equation}
\begin{equation}
\partial _{\nu }U^{\nu \mu }+m^{2}R^{\mu }\psi =0,  \label{eq18-12}
\end{equation}
or equivalently 
\begin{equation}
\left( \square +m^{2}\right) R^{\mu }\psi =0;\ \partial _{\mu }R^{\mu }\psi
=0.  \label{eq18-13}
\end{equation}

So, all elements of the column matrix $R^{\mu }\psi $ are components vector
fields of mass $m$ obeying Proca equation; being the elements of $R^{\mu
\alpha }\psi $ equal to $\frac{-i}{m}$ times the field strength tensor of
the vector field of which the corresponding elements of $R^{\mu }\psi $ are
components. So, similarly to the spin 0 case, this procedure selects the
spin 1 content of DKP theory, making explicitly clear that it describes a
massive vectorial particle.

Moreover, for the spin 0 case, we can choose a 5 degree irreducible
representation of the $\beta ^{\mu }$ matrices in such a way that 
\begin{equation}
P\psi =P\left( 
\begin{array}{c}
\psi _{0} \\ 
\psi _{1} \\ 
\psi _{2} \\ 
\psi _{3} \\ 
\psi _{4}
\end{array}
\right) =\left( 
\begin{array}{c}
0 \\ 
0 \\ 
0 \\ 
0 \\ 
\psi _{4}
\end{array}
\right) ;\ P_{\mu }\psi =\left( 
\begin{array}{c}
0 \\ 
0 \\ 
0 \\ 
0 \\ 
\psi _{\mu }
\end{array}
\right) \ \left( \mu =0,...,3\right) .  \label{eq18b}
\end{equation}

This representation is 
\begin{equation}
\beta ^{0}=\left( 
\begin{array}{ccccc}
0 & 0 & 0 & 0 & 1 \\ 
0 & 0 & 0 & 0 & 0 \\ 
0 & 0 & 0 & 0 & 0 \\ 
0 & 0 & 0 & 0 & 0 \\ 
1 & 0 & 0 & 0 & 0
\end{array}
\right) ;\ \beta ^{1}=\left( 
\begin{array}{ccccc}
0 & 0 & 0 & 0 & 0 \\ 
0 & 0 & 0 & 0 & 1 \\ 
0 & 0 & 0 & 0 & 0 \\ 
0 & 0 & 0 & 0 & 0 \\ 
0 & -1 & 0 & 0 & 0
\end{array}
\right) ;  \label{eq18c}
\end{equation}
\begin{equation}
\beta ^{2}=\left( 
\begin{array}{ccccc}
0 & 0 & 0 & 0 & 0 \\ 
0 & 0 & 0 & 0 & 0 \\ 
0 & 0 & 0 & 0 & 1 \\ 
0 & 0 & 0 & 0 & 0 \\ 
0 & 0 & -1 & 0 & 0
\end{array}
\right) ;\ \beta ^{3}=\left( 
\begin{array}{ccccc}
0 & 0 & 0 & 0 & 0 \\ 
0 & 0 & 0 & 0 & 0 \\ 
0 & 0 & 0 & 0 & 0 \\ 
0 & 0 & 0 & 0 & 1 \\ 
0 & 0 & 0 & -1 & 0
\end{array}
\right) .  \label{eq18d}
\end{equation}

From now on we will use this specific representation. So, equation (\ref
{eq17}) and (\ref{eq18b}) will result in (for $\nu =0,1,2,3$) 
\begin{equation}
\psi _{\nu }=\frac{i}{m}\partial ^{\nu }\left( \psi _{4}\right) .
\label{eq18f}
\end{equation}

We can now make 
\begin{equation}
\psi _{4}=\sqrt{m}\varphi ,  \label{eq18g}
\end{equation}
where $\varphi $ is a scalar field, obtaining 
\begin{equation}
\psi _{\nu }=\frac{i}{\sqrt{m}}\partial _{\nu }\varphi  \label{eq18h}
\end{equation}
so that 
\begin{equation}
\psi =\left( 
\begin{array}{c}
\frac{i}{\sqrt{m}}\partial _{\nu }\varphi \\ 
\sqrt{m}\varphi
\end{array}
\right)  \label{eq19}
\end{equation}
and consequently

\begin{equation}
P\psi =\left( 
\begin{array}{c}
0_{4\times 1} \\ 
\sqrt{m}\varphi
\end{array}
\right) ,\ P^{\mu }\psi =\frac{i}{\sqrt{m}}\left( 
\begin{array}{c}
0_{4\times 1} \\ 
\partial ^{\mu }\varphi
\end{array}
\right) ,\ \square \varphi +m^{2}\varphi =0.  \label{eq20}
\end{equation}

Here we will call expression (\ref{eq19}) for $\psi $ the {\bf physical form}
of DKP field. If we use this representation for the $\beta $ matrices,
together with the physical form of $\psi $, in DKP Lagrangian (\ref{eq6}) we
get 
\begin{equation}
{\cal L}=-\frac{1}{2}\left( \varphi ^{\ast }\square \varphi +\varphi \square
\varphi ^{\ast }\right) -m^{2}\varphi ^{\ast }\varphi ,  \label{eq21}
\end{equation}
which can be integrated by parts giving as result, except by a
four-divergence term, the usual KG Lagrangian 
\begin{equation}
{\cal L}=\partial ^{\mu }\varphi ^{\ast }\partial _{\mu }\varphi
-m^{2}\varphi ^{\ast }\varphi .  \label{eq22}
\end{equation}

Similarly, for the case of spin 1 particles, a 10 degree irreducible
representation of the matrices can be found such that the first four
components of the field $\psi $ are the components of a four vector $B^{\mu
} $, being selected by $R^{\mu }$. The other 6 components of $\psi $ are the
6 independent components of the strength tensor of $B^{\mu }$. These
components are selected by $R^{\mu \nu }$ so that when we apply these
operators to equation (\ref{eq18-12}) it becomes the Proca equation for the
field $B^{\mu }$. Anyway, we will not make use of this specific
representation so we just mention this possibility here for the sake of
completness without working out the complete calculations.

\section{Local gauge invariance and the interaction La\-gran\-gian}

When one interacts DKP field with electromagnetic field, a close attention
must be taken when interpreting the physical content of the theory. After
performing the minimal coupling to the DKP Lagrangian we obtain as an
interaction term 
\begin{equation}
{\cal L}_{I}=eA_{\mu }\overline{\psi }\beta ^{\mu }\psi ,  \label{eq34}
\end{equation}
so that, to regain the interaction term in KG theory, one may be tempted to
use the expression (\ref{eq19}) for $\psi $ in the interaction term above,
getting as result 
\begin{equation}
{\cal L}_{I}=ieA^{\mu }\left( \varphi ^{\ast }\partial _{\mu }\varphi
-\partial _{\mu }\left( \varphi ^{\ast }\right) \varphi \right) ,
\label{eq35}
\end{equation}
that is, a linear term different from the quadratic term obtained from KG
Lagrangian. This result gave rise to the interpretation that DKP and KG
theories were equivalent only in the free field case. Our intention here is
to show that this difference arises from the incorrect use of the physical
form of DKP field (\ref{eq19}) in a situation where it is no longer valid,
since it was obtained {\bf for free fields}, which is not the case anymore.

First of all, we can easily see that expression (\ref{eq19}) for the
physical form is incompatible with gauge invariance. This comes from the
fact that, under local gauge transformations, we must have 
\begin{equation}
\psi \rightarrow \psi 
{\acute{}}%
=e^{ie\alpha \left( x\right) }\psi ;\ \varphi \rightarrow \varphi 
{\acute{}}%
=e^{ie\alpha \left( x\right) }\varphi ,  \label{eq23}
\end{equation}
for DKP and KG fields, respectively. But using the transformation property
of $\varphi $ in the expression (\ref{eq19}) for the physical form results
in 
\begin{equation}
\psi 
{\acute{}}%
=\left( 
\begin{array}{c}
\frac{i}{\sqrt{m}}\partial _{\nu }\varphi 
{\acute{}}%
\\ 
\sqrt{m}\varphi 
{\acute{}}%
\end{array}
\right) =\left( 
\begin{array}{c}
\frac{i}{\sqrt{m}}\left( iee^{ie\alpha \left( x\right) }\varphi \partial
_{\nu }\alpha \left( x\right) +e^{ie\alpha \left( x\right) }\partial _{\nu
}\varphi \right) \\ 
e^{ie\alpha \left( x\right) }\sqrt{m}\varphi
\end{array}
\right) \neq e^{ie\alpha \left( x\right) }\psi ,  \label{eq28}
\end{equation}
so that it becomes obvious that the expression obtained in the free field
case is no longer valid.

The solution to this problem would be simply change the\ physical form of $%
\psi $ from that given by equation (\ref{eq19}) to 
\begin{equation}
\psi =\left( 
\begin{array}{c}
\frac{i}{\sqrt{m}}D_{\nu }\varphi \\ 
\sqrt{m}\varphi
\end{array}
\right) ,  \label{eq29}
\end{equation}
where $D_{\mu }=\partial _{\mu }-ieA_{\mu }$, which obviously transforms as
necessary to keep the compatibility between both transformations in equation
(\ref{eq23}). More than this, we can easily show that this change in the
expression for the physical form arises naturally when we analyse DKP
Lagrangian with minimal coupling, showing explicitly that the direct
application of equation (\ref{eq19}) in this case is a mistake. Indeed,
starting with the DKP Lagrangian (\ref{eq6}) we perform the minimal coupling
obtaining as result 
\begin{equation}
{\cal L}=\frac{i}{2}\left( \overline{\psi }\beta ^{\mu }D_{\mu }\psi -D_{\mu
}^{\ast }\left( \overline{\psi }\right) \beta ^{\mu }\psi \right) -m%
\overline{\psi }\psi ,  \label{eq36}
\end{equation}
or 
\begin{equation}
{\cal L}=\frac{i}{2}\overline{\psi }\beta ^{\mu }\overleftrightarrow{%
\partial }_{\mu }\psi -m\overline{\psi }\psi +e\overline{\psi }A_{\mu }\beta
^{\mu }\psi ,  \label{eq37}
\end{equation}
and from these we obtain the minimally coupled DKP equation

\begin{equation}
\left( i\beta ^{\mu }D_{\mu }-m\right) \psi =0.  \label{eq30}
\end{equation}

Applying the operators $P$ and $P^{\nu }$ we obtain 
\begin{equation}
D_{\nu }\left( P^{\nu }\psi \right) =\frac{m}{i}P\psi  \label{eq31}
\end{equation}
and 
\begin{equation}
P^{\nu }\psi =\frac{i}{m}D^{\nu }\left( P\psi \right) ,  \label{eq32}
\end{equation}
so that 
\begin{equation}
D_{\nu }D^{\nu }\left( P\psi \right) +m^{2}\left( P\psi \right) =0.
\label{eq32b}
\end{equation}

This shows that all elements of the column matrix $P\psi $ are scalar fields
of mass $m$ obeying KG equation with minimal coupling, while the elements of 
$P^{\mu }\psi $ are $\frac{i}{m}$ times the {\bf covariant derivatives} of
the corresponding elements of $P\psi $. Following exactly the same steps and
using the same representation for the matrices $\beta ^{\mu }$ used to
obtain expression (\ref{eq19}) in the free field case, we see that the
correct physical form for $\psi $ when we have electromagnetic interaction
is given by equation (\ref{eq29}). Besides that, we have 
\begin{equation}
P\psi =\left( 
\begin{array}{c}
0_{4\times 1} \\ 
\sqrt{m}\varphi
\end{array}
\right) ,\ P^{\mu }\psi =\frac{i}{\sqrt{m}}\left( 
\begin{array}{c}
0_{4\times 1} \\ 
D^{\mu }\varphi
\end{array}
\right) ,\ D^{\mu }D_{\mu }\varphi +m^{2}\varphi =0.  \label{eq33}
\end{equation}

Furthermore, to regain KG interaction Lagrangian it is not enough to use the
expression (\ref{eq29}) for $\psi $ in the DKP interaction term (\ref{eq34}%
), we must consider the whole DKP Lagrangian since now $\psi $ has covariant
derivatives in its components and, consequently, ``spread'' the interaction
terms throughout the whole DKP Lagrangian. So, we must substitute expression
(\ref{eq29}) in the complete minimally coupled DKP Lagrangian given by
expression (\ref{eq36}) or (\ref{eq37}). This results in 
\begin{equation}
{\cal L}=-\frac{1}{2}\left( \varphi ^{\ast }\partial ^{\mu }D_{\mu }\varphi
+\varphi \partial ^{\mu }D_{\mu }^{\ast }\varphi ^{\ast }\right) +ie\frac{1}{%
2}A^{\mu }\left( \varphi ^{\ast }\partial _{\mu }\varphi -\varphi \partial
_{\mu }\varphi ^{\ast }\right) +e^{2}A^{\mu }A_{\mu }\varphi \varphi ^{\ast
}-m^{2}\varphi ^{\ast }\varphi ,  \label{eq38}
\end{equation}
\begin{equation}
{\cal L}=\frac{1}{2}\left( \partial ^{\mu }\varphi ^{\ast }D_{\mu }\varphi
+\partial ^{\mu }\varphi D_{\mu }^{\ast }\varphi ^{\ast }\right) +ie\frac{1}{%
2}A^{\mu }\left( \varphi ^{\ast }\partial _{\mu }\varphi -\varphi \partial
_{\mu }\varphi ^{\ast }\right) +e^{2}A^{\mu }A_{\mu }\varphi \varphi ^{\ast
}-m^{2}\varphi ^{\ast }\varphi ,  \label{eq39}
\end{equation}
\begin{equation}
{\cal L}=\partial ^{\mu }\varphi ^{\ast }\partial _{\mu }\varphi +ieA^{\mu
}\left( \varphi ^{\ast }\partial _{\mu }\varphi -\varphi \partial _{\mu
}\varphi ^{\ast }\right) +e^{2}A^{\mu }A_{\mu }\varphi \varphi ^{\ast
}-m^{2}\varphi ^{\ast }\varphi ,
\end{equation}
where an integration by parts was performed in the first term of equation (%
\ref{eq38}). This is exactly the usual KG Lagrangian with minimal coupling
so that we obtain the correct interaction term 
\begin{equation}
{\cal L}_{I}=ieA^{\mu }\left( \varphi ^{\ast }\partial _{\mu }\varphi
-\varphi \partial _{\mu }\varphi ^{\ast }\right) +e^{2}A^{\mu }A_{\mu
}\varphi \varphi ^{\ast }.  \label{eq41}
\end{equation}

Thus we see that, using the correct physical form for the DKP field $\psi $
in the minimal coupling case, we can recover the KG Lagrangian with the
correct minimal coupling interaction term. Consequently, the equivalence of
these theories is kept when electromagnetic interaction comes to play. The
apparent difference reported in literature comes from the use of a physical
form for the DKP field, obtained in the free case, that is no longer valid
with the presence of the electromagnetic field and is incompatible with the
requirement of local gauge invariance.

\section{The physical components and the anomalous terms}

Now we will analyse the presence of an apparently anomalous term lacking
physical interpretation in Hamiltonian and second order forms of minimally
coupled DKP theory. These forms are obtained starting from the DKP equation
with minimal coupling, equation (\ref{eq30}), and contracting from the left
with $D_{\alpha }\beta ^{\alpha }\beta ^{\nu }$, which results in 
\begin{equation}
i\beta ^{\alpha }\beta ^{\nu }\beta ^{\mu }D_{\alpha }D_{\mu }\psi
-mD_{\alpha }\beta ^{\alpha }\beta ^{\nu }\psi =0.  \label{eq43}
\end{equation}

After some algebraic calculation the above expression reduces to 
\begin{equation}
D^{\nu }\psi =\beta ^{\mu }\beta ^{\nu }D_{\mu }\psi +\frac{e}{2m}F_{\alpha
\mu }\left( \beta ^{\mu }\beta ^{\nu }\beta ^{\alpha }+\beta ^{\mu }\eta
^{\nu \alpha }\right) \psi ,  \label{eq44}
\end{equation}
where the relation $\left[ D_{\mu },D_{\nu }\right] =-ieF_{\mu \nu }$ was
used. Making $\nu =0$ in this last expression we have 
\begin{equation}
D^{0}\psi -\left( \beta ^{0}\right) ^{2}D_{0}\psi -\beta ^{i}\beta
^{0}D_{i}\psi -\frac{e}{2m}F_{\alpha \mu }\left( \beta ^{\mu }\beta
^{0}\beta ^{\alpha }+\beta ^{\mu }\eta ^{0\alpha }\right) \psi =0,
\label{eq45}
\end{equation}
while multiplying equation (\ref{eq30}) from left by $-i\beta ^{0}$ results
in 
\begin{equation}
\left( \beta ^{0}\right) ^{2}D_{0}\psi +\beta ^{0}\beta ^{i}D_{i}\psi
+im\beta ^{0}\psi =0.  \label{eq46}
\end{equation}

Then, summing these equations we get the Hamiltonian (Schr\"{o}dinger like)
form of DKP wave equation 
\begin{equation}
i\partial _{t}\psi =H\psi ,  \label{eq47}
\end{equation}
where 
\begin{equation}
H=i\left[ \beta ^{i},\beta ^{0}\right] D_{i}+i\frac{e}{2m}F_{\alpha \mu
}\left( \beta ^{\mu }\beta ^{0}\beta ^{\alpha }+\beta ^{\mu }\eta ^{0\alpha
}\right) -eA^{0}+m\beta ^{0}.  \label{eq48}
\end{equation}

Finally, contracting equation (\ref{eq44}) with $D_{\nu }$ we get as result
a second order wave equation 
\begin{equation}
D_{\nu }D^{\nu }\psi +m^{2}\psi -\frac{i}{2}eF_{\mu \nu }S^{\mu }{}^{\nu
}\psi -\frac{e}{2m}\left( \beta ^{\mu }\beta ^{\nu }\beta ^{\alpha }+\beta
^{\mu }\eta ^{\nu \alpha }\right) D_{\nu }\left( F_{\alpha \mu }\psi \right)
=0.  \label{eq49}
\end{equation}

The anomalous term is the one proportional to $e/2m$ in equations (\ref{eq44}%
) and (\ref{eq49}) and in Hamiltonian (\ref{eq48}). This denomination is due
to the fact, already noticed by Kemmer in his original work \cite{Kemmer},
that it has no apparent physical interpretation, contrary to the others
terms in these equations, which have physical interpretations analogous to
similar terms obtained when working with Dirac equation.

Recently Nowakowski \cite{Nowakowski} settled the question by showing that
the above mentioned term has no physical meaning and simply disapears when
one works with the physical components of the DKP field\footnote{%
Moreover, he shows that the solutions of the second order equation (\ref
{eq49}) will not always be solutions of the first order DKP equation. So, it
is not a good analog to Dirac's second order equation. He also shows that (%
\ref{eq49}) is just one among a class of second order equations that can be
obtained from DKP equation.}. This is done by the use a specific choice of
the $10\times 10$ and $5\times 5$ representations of DKP algebra. Our
intention here is to show that this result can be easily obtained in any
representation of the algebra (it is not even necessary to be an irreducible
one) through the use of the operators, defined in Section 2, that select the
spin 0 and spin 1 sectors of the theory from any choice of the $\beta $
matrices satisfying DKP algebra.

First, for the case of the spin 0 sector of the theory, we can apply the
operator $P$ defined by equation (\ref{eq11}) to the second order equation (%
\ref{eq49}) obtaining 
\begin{equation}
D_{\nu }D^{\nu }\left( P\psi \right) +m^{2}\left( P\psi \right) -\frac{i}{2}%
eF_{\mu \nu }PS^{\mu }{}^{\nu }\psi -\frac{e}{2m}P\left( \beta ^{\mu }\beta
^{\nu }\beta ^{\alpha }+\beta ^{\mu }\eta ^{\nu \alpha }\right) D_{\nu
}\left( F_{\alpha \mu }\psi \right) =0  \label{eq50}
\end{equation}
\begin{equation}
D_{\nu }D^{\nu }\left( P\psi \right) +m^{2}\left( P\psi \right) -\frac{e}{2m}%
\left( \eta ^{\mu \nu }P^{\alpha }+P^{\mu }\eta ^{\nu \alpha }\right) D_{\nu
}\left( F_{\alpha \mu }\psi \right) =0  \label{eq51}
\end{equation}
\begin{equation}
D_{\nu }D^{\nu }\left( P\psi \right) +m^{2}\left( P\psi \right) =0,
\label{eq52}
\end{equation}
where we have used equations (\ref{eq12}) and (\ref{eq13}). So we see that
when we select the spin 0 sector $P\psi $ the anomalous term vanishes. More
than this, the spin-field interaction term \cite{Kemmer} $F_{\mu \nu }S^{\mu
}{}^{\nu }$ also vanishes, as it should be for a scalar field. When applying 
$P^{\lambda }$ to this equation we get as result the covariant derivative $%
D^{\lambda }$ of KG equation (\ref{eq52}), but we must remember that the
relations between $P$ and $P^{\lambda }$ are now given by equations (\ref
{eq31}) and (\ref{eq32}) and not by equations (\ref{eq16}) and (\ref{eq17})
obtained for the free field case. Analogously, when using $P$ and $%
P^{\lambda }$ on equations (\ref{eq44}) and (\ref{eq47}) we obtain trivial
equalities or regain KG equation (\ref{eq52}).

Now, in order to analyse the spin 1 sector, we must derive the relation
between $R^{\mu }\psi $ and $R^{\mu \nu }\psi $ in the presence of minimal
coupling, in the same way that we have done in the previous section with the
operators $P$ and $P^{\mu }$. So, we apply those operators to the DKP
equation with minimal coupling, equation (\ref{eq30}), obtaining 
\begin{equation}
D_{\alpha }\left( R^{\mu \alpha }\psi \right) =\frac{m}{i}\left( R^{\mu
}\psi \right) ,  \label{eq53}
\end{equation}
and 
\begin{equation}
\left( R^{\mu \nu }\psi \right) =-\frac{i}{m}U^{I\mu \nu },  \label{eq54}
\end{equation}
where 
\begin{equation}
U^{I\mu \nu }=D^{\mu }R^{\nu }\psi -D^{\nu }R^{\mu }\psi ,  \label{eq55}
\end{equation}
is the covariant stress tensor of the massive vector field $R^{\mu }\psi $
interacting minimally with electromagnetic field. This tensor can be written
explicitly as 
\begin{equation}
U^{I\mu \nu }=U^{\mu \nu }-ie\left( A^{\mu }R^{\nu }\psi -A^{\nu }R^{\mu
}\psi \right) .  \label{eq56}
\end{equation}
It is interesting to mention that combining these results we have the
minimally coupled Proca equation 
\begin{equation}
D_{\alpha }\left( U^{I\alpha \mu }\right) +m^{2}\left( R^{\mu }\psi \right)
=0.  \label{eq57}
\end{equation}

Now, when we apply the operator $R^{\lambda }$ to equation (\ref{eq49}) we
get 
\begin{eqnarray}
D_{\nu }D^{\nu }R^{\lambda }\psi +m^{2}R^{\lambda }\psi &-&\frac{i}{2}%
eF_{\mu \nu }R^{\lambda }S^{\mu \nu }\psi  \nonumber \\
&-&\frac{e}{2m}\left( R^{\lambda \mu }\beta ^{\nu }\beta ^{\alpha
}+R^{\lambda \mu }\eta ^{\nu \alpha }\right) D_{\nu }\left( F_{\alpha \mu
}\psi \right) =0,  \label{eq58}
\end{eqnarray}
which can be shown to reduce, after some algebraic manipulation, to the
minimally coupled Proca equation (\ref{eq57}). When we apply the operator $%
R^{\lambda \alpha }$ to the second order equation (\ref{eq49}) we simply
regain the $U^{I\mu \nu }$ definition, equation (\ref{eq56}).

Similar results follow to the others equations: we regain definitions or get
trivial identities. As example, applying $R^{\lambda }$ to equation (\ref
{eq44}) will result in 
\begin{equation}
D_{\mu }R^{\mu }\psi =\frac{ie}{2m^{2}}F_{\alpha \mu }U^{I\mu \alpha },
\label{eq59}
\end{equation}
where $F_{\alpha \mu }$ is the usual stress tensor for electromagnetic field 
$A_{\mu }$. But this result can be obtained directly from the covariant
derivative of Proca equation (\ref{eq57}).

The important result is that when we select the physical components of DKP
field $\psi $ the anomalous term is eliminated, so it has no physical
meaning. And, as mentioned before, this result {\bf is not} dependent upon a
specific choice of the $\beta $ matrices nor the degree of the
representation, being quite general.

\section{Conclusions and comments}

In this work we discussed two points relative to the minimal electromagnetic
coupling in DKP theory that is not usually understood in literature
correctly.

First we showed that a reported difference between the interaction term for
scalar bosons in spin 0 DKP theory and KG theory does not exist. This
apparent difference was showed to be caused by the incorrect use of an
expression for the physical form of DKP field that is valid only in the free
field case, not in the minimally coupled case. When the correct physical
form of $\psi $ in the presence of electromagnetic interaction is used, we
find no difference between the interaction terms in both theories. We have
also shown that the gauge invariance principle shows clearly the origin of
the problem since the free field physical form of $\psi $ is incompatible
with gauge invariance: changing the expression for physical form is
necessary to keep gauge invariance. Moreover, the correct expression was
easily obtained using DKP Lagrangian with minimal coupling, the minimally
coupled DKP equation, the projectors of the physical components and a
convenient choice of $\beta $ matrices (the same used in the free field
case).

In addition we showed that the apparently anomalous term in second order and
Hamiltonian forms of DKP equation disappears when we select the physical
components of DKP field, so that this term has no physical meaning. This
conclusion is exactly the same presented by Nowakowski \cite{Nowakowski}
but, it seems to us, was obtained here in a more clear and becoming way.
Furthermore, the use of the operators that select the spin 0 and 1 sectors
of the theory do not demand the choice of an specific representation of the $%
\beta $ matrices. Finally, it is also interesting to notice that, when we
project the physical components of DKP field, the second order equation (\ref
{eq49}) is reduced to the well known second order equations for scalar and
vector fields, i.e. Klein-Gordon and Proca equations.

\section{Acknowlegdments}

J.T.L. and B.M.P. would like to thank CAPES's PICDT program and CNPq,
respectively, for partial support. R.G.T. and J.S.V. thank CAPES and FAPESP
(grant 98/00268-1), respectively, for full support.

\end{document}